\documentclass[a4paper,11pt]{article}
\usepackage{pos}

\usepackage[symbol]{footmisc}
\usepackage[english]{babel}
\usepackage{graphicx}
\usepackage{graphics}
\usepackage{braket}
\usepackage{bbold}
\usepackage{amsmath}
\usepackage{nicefrac}
\usepackage{dcolumn}
\usepackage{bm}
\usepackage{slashed}
\usepackage{datetime}
\usepackage{mciteplus}
\usepackage{multirow}
\usepackage{siunitx}
\usepackage{booktabs}
\usepackage{color, soul}
\usepackage[usenames,dvipsnames]{xcolor}
\usepackage{float}
\usepackage[utf8]{inputenc}
\usepackage[normalem]{ulem}
\usepackage{mathtools}
\usepackage{setspace}
\usepackage{comment}
\renewcommand{\thefootnote}{\fnsymbol{footnote}}

\newcommand{\drv}{{\rm d}}

\newcommand{\LQCD}{\Lambda_{\rm QCD}}

\newcommand{\Jps}{J/\psi}
\newcommand{\Yps}{\Upsilon}
\newcommand{\ecs}{\eta_c}
\newcommand{\ebs}{\eta_b}

\newcommand{{\HFNRevo}}{\tt HF-NRevo}

\title{Collinear fragmentation of pseudoscalar quarkonia from NLO NRQCD}
\ShortTitle{Collinear fragmentation of pseudoscalar quarkonia from NLO NRQCD}

\author[a]{Francesco Giovanni Celiberto}
\author*[a]{Francesca Lonigro}

\affiliation[a]{Departamento de Física y Matemáticas, Universidad de Alcalá (UAH), Campus Universitario, \\ Alcalá de Henares, E-28805, Madrid, Spain}

\emailAdd{francesco.celiberto@uah.es}
\emailAdd{francesca.lonigro@uah.es}

\abstract{We discuss a new family of publicly available collinear fragmentation functions (FFs) for pseudoscalar quarkonia, the {\tt NRFF1.0} set.
It builds upon the Heavy-Flavor Non-Relativistic evolution (HF-NRevo) scheme, designed to describe heavy-hadron formation through leading-power fragmentation at moderate and large transverse momentum.
Heavy-quarkonium production naturally involves both perturbative and non-perturbative QCD dynamics, from the production of the heavy $Q\bar{Q}$ pair to its formation into a physical bound state.
Within {\tt NRFF1.0}, Non-Relativistic Quantum Chromodynamics (NRQCD) provides the theoretical framework for calculating the FF inputs at the initial scale, which are subsequently evolved through the HF-NRevo scheme.
This setup provides a precision baseline for investigating the underlying partonic hierarchy and jet structure across the moderate- to high-transverse-momentum regime.
The explicit treatment of partonic channels and heavy-flavor thresholds makes this framework particularly suitable for exploring quarkonium-in-jet fragmentation, jet-quenching sensitivity, energy-loss mechanisms, and the emergence of medium-modified fragmentation patterns in the quark-gluon plasma.}

\FullConference{The 33rd International Workshop on Deep Inelastic Scattering and Related Subjects (DIS2026)\\
4-8 May 2026\\
Bologna, Italy\\}

\begin{document}
\maketitle

\section{Introduction}
\label{sec:introduction}
Hadrons featuring either open or hidden heavy flavors act as crucial tools for investigating fundamental interactions. 
Because heavy quarks are expected to couple with particles beyond the Standard Model, they play a pivotal role in the search for new physics. 
Additionally, since their masses exceed the QCD confinement scale, they provide an ideal laboratory for examining the perturbative regime of the strong interaction.
Often described as the ``hydrogen atoms’’ of QCD~\cite{Pineda:2011dg}, quarkonium states provide profound insights into the mechanisms of hadronization, serving as a link between high-precision perturbative QCD and the exploration of the internal proton structure. 
For instance, hadronic decays of $S$-wave bottomonium states enable accurate determinations of $\alpha_s$~\cite{Brambilla:2007cz,Proceedings:2019pra}, whereas forward quarkonium production places valuable constraints on the positivity of gluon PDFs in the low-$x$ and low-$Q^2$ regions~\cite{Altarelli:1998gn,Candido:2020yat}. Furthermore, quarkonia provide sensitive probes of the three-dimensional structure of the proton at both small~$x$~\cite{Hentschinski:2020yfm,Celiberto:2018muu,Bolognino:2018rhb,Bolognino:2021niq,Celiberto:2019slj,Silvetti:2022hyc,Kang:2023doo} and intermediate $x$ scales~\cite{Boer:2015pni,Lansberg:2017dzg,Bacchetta:2020vty,Bacchetta:2024fci,Celiberto:2021zww}. 
Additionally, inclusive photoproduction of a $\Jps$ meson alongside a charm jet at the future Electron-Ion Collider (EIC) is expected to offer a direct probe of intrinsic-charm components in the proton~\cite{NNPDF:2023tyk,Flore:2020jau}.
Understanding quarkonium hadronization continues to represent a major theoretical challenge, as no single description can comprehensively account for all experimental observations. 
The effective field theory of NRQCD provides a systematic framework to address this problem~\cite{Caswell:1985ui,Bodwin:1994jh}. 
Within this framework, physical quarkonia are described in terms of distinct Fock states, organized through an expansion in both $\alpha_s$ and the relative velocity $v$ of the constituent $[Q\bar Q]$ pair. 
Production cross sections are consequently factorized into perturbatively calculable Short-Distance Coefficients (SDCs) and nonperturbative Long-Distance Matrix Elements (LDMEs).
At moderate and large transverse momenta, quarkonium production can be efficiently described within a leading-power fragmentation picture, in which the short-distance production of a parton is separated from its subsequent transition into the observed bound state. 
This motivates the construction of dedicated quarkonium FF sets. 
The sets considered in this work are built within the HF-NRevo framework~\cite{Celiberto:2025euy,Celiberto:2024mex_article,Celiberto:2024bxu,Celiberto:2024rxa,Celiberto:2025xvy,Celiberto:2026rzi,Celiberto:2026zss}. 
This approach incorporates next-to-leading-order (NLO) NRQCD inputs at the initial scale, performs their evolution through the DGLAP equations, and accounts for missing higher-order uncertainties through a replica-based Monte Carlo technique~\cite{Forte:2002fg}.

\section{Quarkonium fragmentation from HF-NRevo}
\label{sec:HFNrevo}

\begin{figure*}[!t]
\centering

   \includegraphics[scale=0.38,clip]{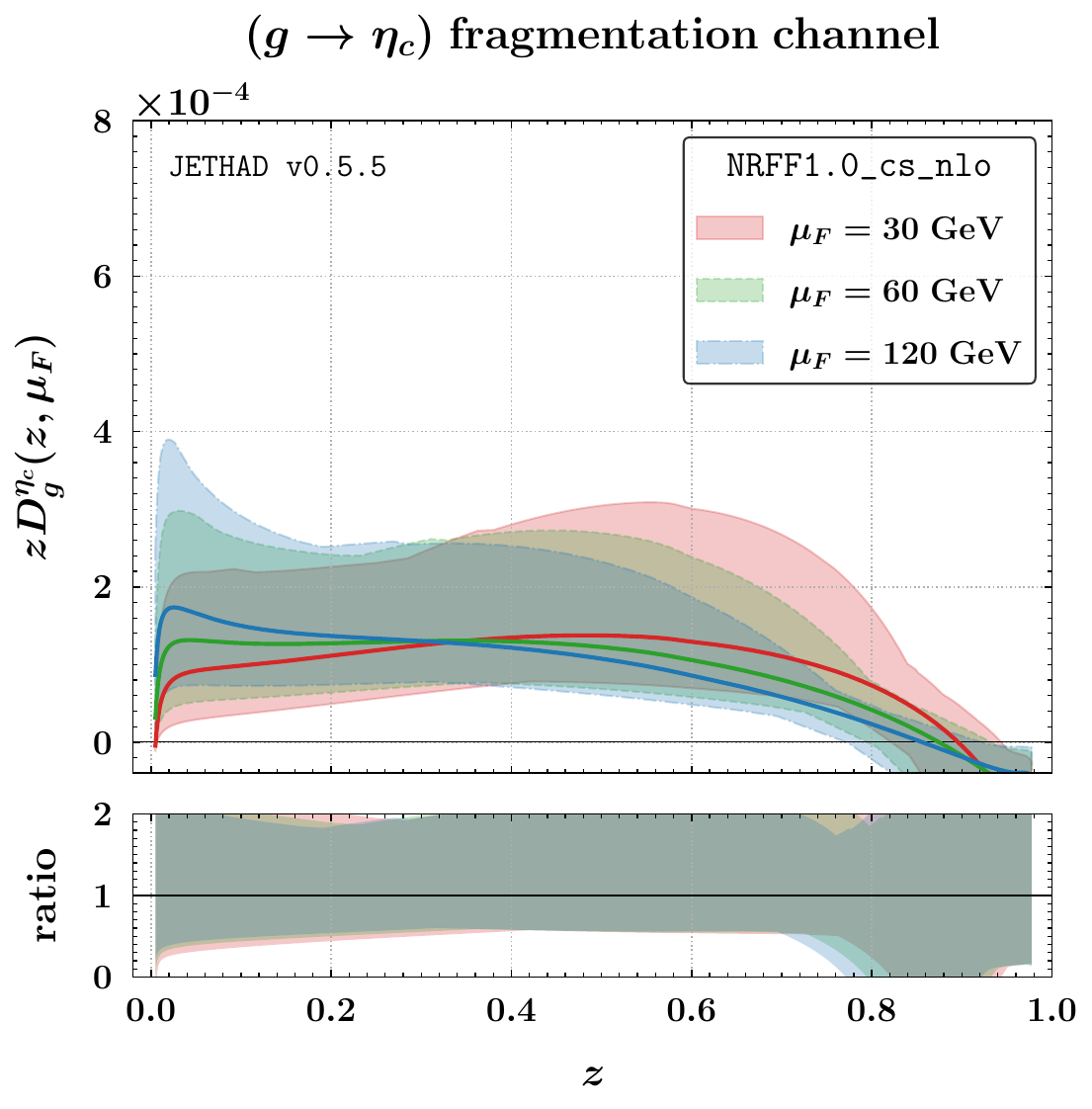}
   \hspace{0.15cm}
   \includegraphics[scale=0.38,clip]{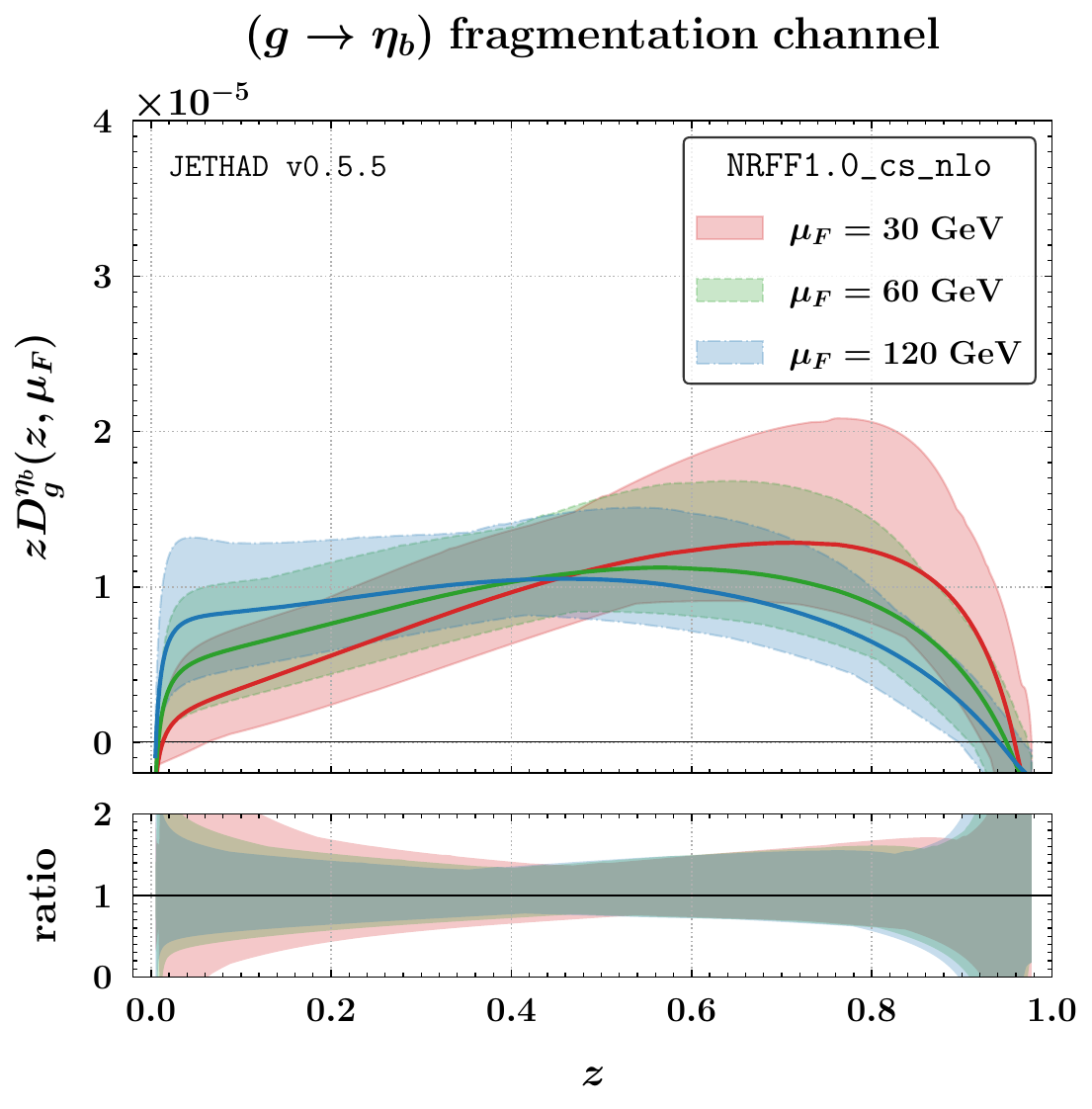}

\caption{NLO gluon FF into color-singlet $\ecs$ ($\ebs$) states.
Figure adapted from~\protect\cite{Celiberto:2025euy}.
}

\label{fig:FFs_bottom}
\end{figure*}

\begin{figure*}[!t]
\centering

   \includegraphics[scale=0.38,clip]{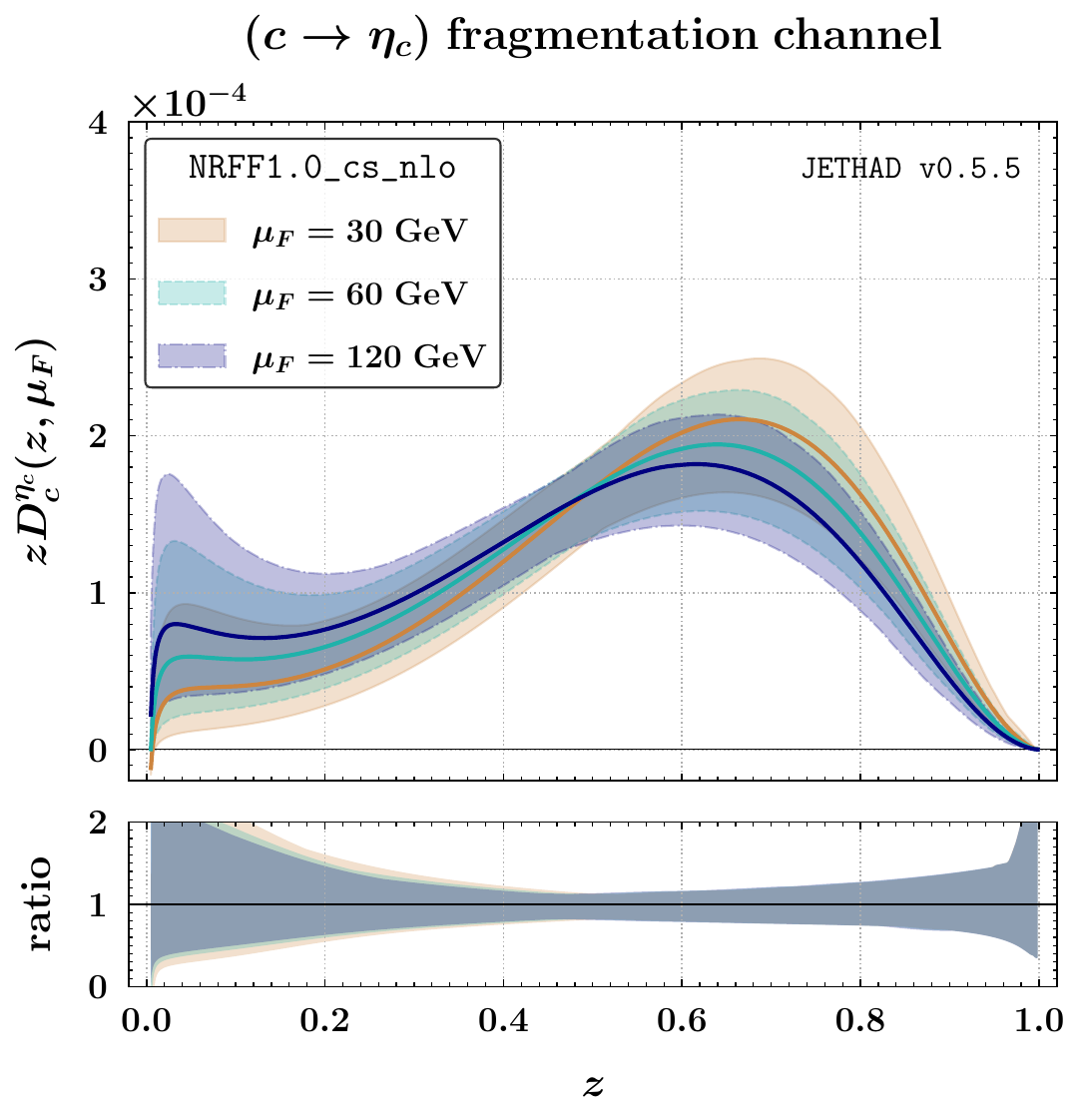}
   \hspace{0.25cm}
   \includegraphics[scale=0.38,clip]{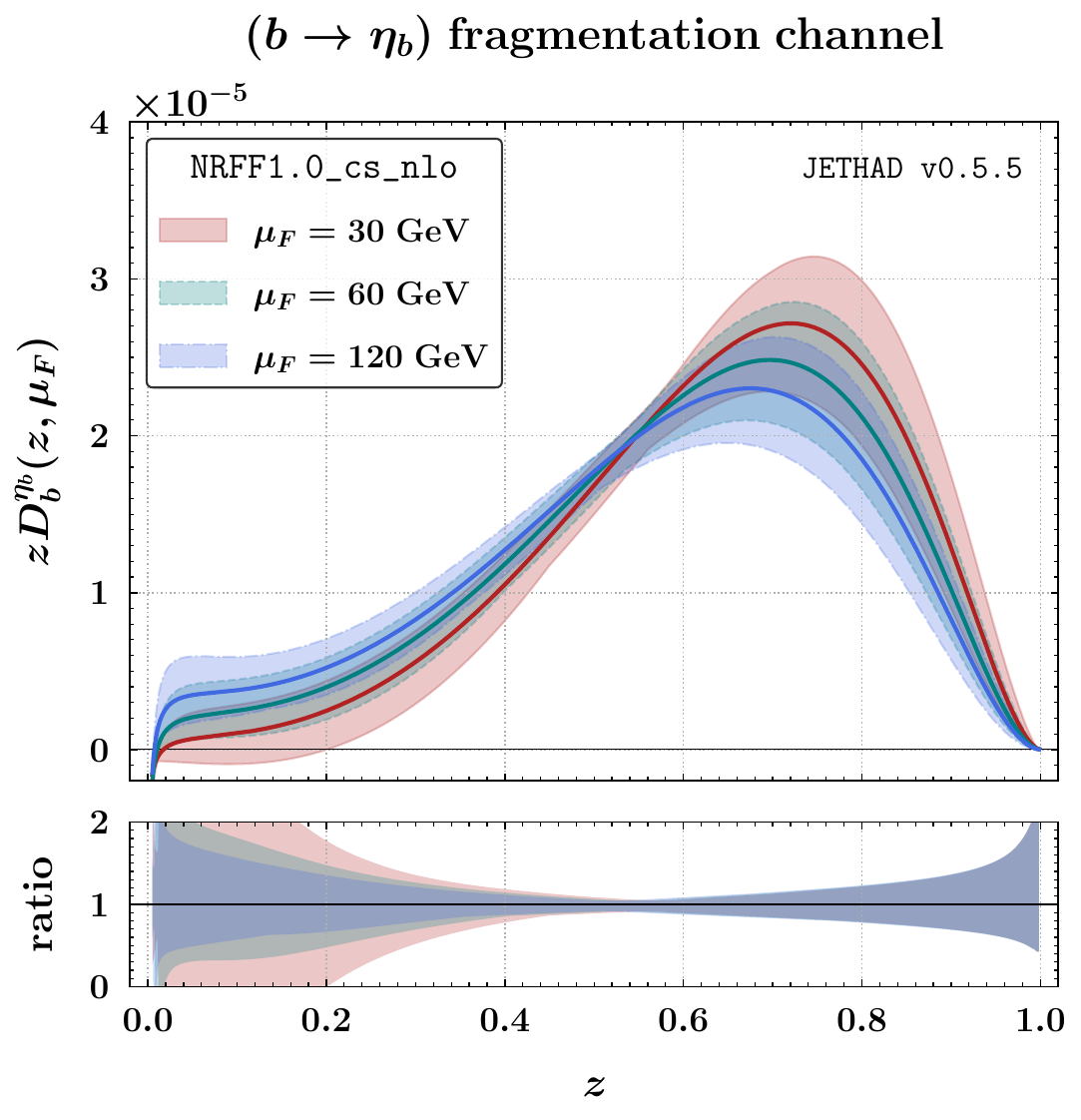}

\caption{NLO charm (bottom) FF into color-singlet $\Jps$ ($\Yps$) states.
Figure adapted from~\protect\cite{Celiberto:2025euy}.
}

\label{fig:FFs_Q}
\end{figure*}
Because the masses of the constituent heavy quarks significantly exceed the characteristic QCD confinement scale, $\LQCD$, the initial conditions for FFs retain a distinctive perturbative character.
This feature enables a consistent implementation within collinear factorization.
To this end, we employ the HF-NRevo framework, which provides a systematic methodology for defining, evolving, and assessing uncertainties in heavy-quarkonium FFs.
The architecture of HF-NRevo rests upon three primary pillars: physical interpretation, scale evolution, and the quantification of systematic uncertainties.
Concerning the first pillar, heavy-quarkonium fragmentation is initially described through a two-parton mechanism formulated within a Fixed-Flavor Number Scheme (FFNS).
This setup provides a natural starting point for subsequent matching onto a Variable-Flavor Number Scheme (VFNS).
The underlying interpretation is further supported by studies of transverse-momentum-dependent (TMD) observables, which reveal complementary singularity structures in shape functions~\cite{Echevarria:2019ynx} and collinear FFs~\cite{Boer:2023zit}.
Within HF-NRevo, the evolution of quarkonium FFs is organized into two sequential phases.
The first is a symbolic stage, denoted as {\tt EDevo}, which performs an expanded and decoupled DGLAP evolution designed to consistently handle heavy-flavor thresholds.
This phase is carried out through the symbolic engine integrated into the {\tt JETHAD} environment~\cite{Celiberto:2020wpk,Celiberto:2022rfj,Celiberto:2022kxx,Celiberto:2020tmb,Bolognino:2021mrc,Celiberto:2021dzy,Celiberto:2021fdp,Celiberto:2022zdg,Celiberto:2022gji,Celiberto:2023fzz,Celiberto:2024mrq,Celiberto:2024swu,Celiberto:2026ooh,Celiberto:2025csa}.
The second phase, {\tt AOevo}, performs the numerical all-order DGLAP evolution.
The third cornerstone of HF-NRevo concerns the assessment of missing higher-order uncertainties (MHOUs) associated with the treatment of heavy-flavor thresholds during evolution.
Specifically, we employ a scale-variation procedure to quantify the sensitivity of the FFs to the renormalization and factorization scales entering the initial-scale inputs.

\section{Quarkonium-in-jet fragmentation}
\label{sec:onium_in_jet}
In recent years, jet substructure observables have become established as highly effective tools for probing the core dynamics of the strong interaction. 
Beyond advancing our fundamental understanding of QCD, these observables open up promising new pathways for identifying signatures of physics beyond the Standard Model. 
In particular, a comprehensive examination of the internal topology of jets—especially those featuring heavy-flavored hadrons—provides invaluable insights into both the perturbative and nonperturbative regimes of QCD~\cite{Procura:2009vm,Bauer:2013bza,Chien:2015ctp,Maltoni:2016ays,Kang:2017glf,Metodiev:2018ftz,Marzani:2019hun,Kasieczka:2020nyd,Nachman:2022emq,Dhani:2024gtx}.
Prominent among these observables are measurements that depend directly on identifying a specific hadron enclosed within the jet cone. 
In the framework of collinear factorization, this production mechanism is theoretically characterized by the Semi-Inclusive Fragmenting Jet Function (SIFJF) formalism. 
At leading power, the SIFJF tracking a parton $i$ that fragments into a designated quarkonium state ${\cal H}_Q$ inside a jet is expressed as~\cite{Kang:2017yde}
\begin{equation}
 \label{eq:SIFJF}
 {\cal F}_i^{\cal H}(z, z_{\cal H}, \mu_F, {\cal R}_{\cal J}) \, = \,
 \sum_{j=q,\bar{q},g} \int_{z_{\cal H}}^1 \frac{\drv \zeta}{\zeta} \,
 {\cal S}(z, z_{\cal H}/\zeta, \mu_F, {\cal R}_{\cal J}) \,
 D_j^{\cal H}(\zeta, \mu_F) \;,
\end{equation}
where $D_j^{\cal H}(\zeta, \mu_F)$ represents the standard $[j \to {\cal H}_Q]$ FF channel, and ${\cal S}(z, z_{\cal H}/\zeta, \mu_F, {\cal R}_{\cal J})$ denotes the perturbatively calculable fragmenting jet coefficients~\cite{Baumgart:2014upa}, which have been determined at NLO accuracy for both the anti-$\kappa_T$ and cone jet clustering algorithms~\cite{Kang:2016ehg}. 
In Eq.~\ref{eq:SIFJF}, the kinematic variable $z$ defines the ratio of the light-cone momentum of the final jet to that of the parent parton $i$. 
Correspondingly, $z_{\cal H}$ denotes the fraction of the jet's light-cone momentum carried by the identified internal hadron. 
Lastly, the geometric parameter ${\cal R}_{\cal J}$ characterizes the radius of the jet.

\section{Bridging to heavy ions}
\label{sec:heavy_ions}
Quarkonium production in heavy-ion collisions is strongly affected by the formation of a deconfined QCD medium.
In particular, a pronounced suppression of charmonium and bottomonium yields is observed relative to baseline measurements in proton–proton collisions.
This suppression is driven by several competing physical mechanisms~\cite{Vogt:1999cu,Satz:2000bn}.
One of the most prominent is \emph{color screening}, whereby thermal color charges in the quark-gluon plasma (QGP) screen the binding interaction between the heavy quark and antiquark~\cite{Shuryak:1980tp,Heinz:2000bk,Braun-Munzinger:2015hba}.
As a consequence, more weakly bound states dissociate more readily, giving rise to a sequential-melting pattern correlated with their binding energies.
For instance, ground-state quarkonia such as the $J/\psi$ and $\Upsilon$ are generally more resilient against dissociation than their radially excited counterparts, $\psi(2S)$ and $\Upsilon(2S)$~\cite{Matsui:1986dk,Grandchamp:2003uw,GayDucati:2003xa}.
Another relevant mechanism is gluon-induced \emph{dissociation}, in which inelastic scattering between thermal gluons and quarkonia leads to the breakup of the bound states.
At higher center-of-mass energies, suppression can be partially counterbalanced by quarkonium \emph{regeneration}, arising from the recombination of heavy quark-antiquark pairs present in the QGP.
The observed yield therefore results from the interplay among suppression and regeneration mechanisms, whose relative importance depends on properties such as the local temperature, medium lifetime, and heavy-quark density~\cite{Andronic:2008gu,Grandchamp:2003uw}.
These phenomena are commonly investigated through transport approaches and hydrodynamic descriptions supplemented by information on in-medium quarkonium properties.
Nevertheless, substantial theoretical uncertainties remain, particularly in the description of quarkonium production at large transverse momentum and for pseudoscalar states such as the $\eta_{c,b}$.

Within this context, the HF-NRevo framework provides a controlled vacuum baseline for future studies of medium-modified quarkonium fragmentation.
Its scale-dependent evolution and explicit treatment of heavy-flavor thresholds offer a natural starting point for incorporating medium-induced effects, including parton energy loss and modifications of fragmentation patterns in dense nuclear matter.
Such developments could eventually be interfaced with complementary descriptions of in-medium hadronization and recombination mechanisms.
More generally, comparing vacuum and medium-modified fragmentation patterns could provide new observables sensitive to the interaction of heavy quarkonia with the QGP.
In this direction, one may investigate apparent distortions of the FF shape induced by effects such as thermal broadening, dissociation, or modified formation dynamics.
Such signatures could be explored through quarkonium-tagged jet measurements at large transverse momentum at the HL-LHC and future high-energy colliders, providing complementary tomographic information on the properties of the QGP.

\section{Conclusions and Outlook}
\label{sec:conclusions}
In this work, we have presented the HF-NRevo framework as a systematic strategy for determining quarkonium FFs within collinear factorization.
Building on this approach, we have introduced the first release of the {\tt NRFF1.0} distribution sets~\cite{Celiberto:2025euy}.
These sets incorporate color-singlet initial-scale inputs for all relevant partonic channels, derived from NLO calculations within NRQCD.
Their subsequent DGLAP evolution consistently accounts for heavy-flavor thresholds, while the associated theoretical uncertainties are quantified through a Monte Carlo replica-based technique designed to estimate missing higher-order effects.
The resulting {\tt NRFF1.0} sets~\cite{Celiberto:2025euy} provide an updated and uncertainty-aware baseline with respect to earlier parametrizations, such as the {\tt ZCW19}$^+$ and {\tt ZCFW22} sets, which have been employed in phenomenological studies of vector quarkonia~\cite{Celiberto:2022dyf,Celiberto:2023fzz} and $B_c$ mesons~\cite{Celiberto:2022keu,Celiberto:2024omj}.

\section*{Acknowledgments}
\label{sec:acknowledgments}

We are supported by the Atracci\'on de Talento Grant n. 2022-T1/TIC-24176 (Madrid, Spain).

\vspace{-0.05cm}
\begingroup
\setstretch{0.6}
\bibliographystyle{bibstyle}
\bibliography{bibliography}
\endgroup

\end{document}